\documentstyle[epsfig]{mn}
\title{Amplitude variability or close frequencies in pulsating stars? The $\delta$ Scuti star FG Vir}

\author[M. Breger and A. A. Pamyatnykh]
{M.~Breger$^1$ and A. A. Pamyatnykh$^{1,2,3}$\\
$^1$ Astronomisches Institut der Universit\"at Wien, T\"urkenschanzstr. 17, A--1180 Wien, Austria\\
$^2$ Copernicus Astronomical Center, Bartycka 18, 00-716 Warsaw, Poland\\
$^3$ Institute of Astronomy, Russian Academy of Sciences, Pyatnitskaya Str. 48,
109017 Moscow, Russia}

\date{Accepted 2005 month day.
      Received 2005 month day;
      in original form 2005 month date}

\begin{document}
\maketitle

\begin{abstract}

The nature of the observed amplitude variability of several modes in the $\delta$~Scuti star FG~Vir
is examined. This is made possible by the extensive photometry obtained during 2002, 2003 and 2004,
as well as the long photometric time base starting in 1992. In this star, three frequencies show
strong amplitude and phase variations. In the power spectrum, these frequencies also show up as
frequency doublets. However, since true amplitude variability of a single frequency
can also lead to (false) frequency doublets in the power spectrum, a specific test examining in detail the observed
amplitude and phase variations of an assumed single frequency is applied. For the frequencies at
12.15 and 23.40~cycle d$^{-1}$ it is shown that amplitude variability
of a single mode can be ruled out. In particular, an important property of beating between two modes is fulfilled:
the amplitude and phase vary synchronously with a phase shift close to 90$\degr$.
The origin of the amplitude variability of a third mode, viz. near 19.86~cycle d$^{-1}$, is not
clear due to the long beat period of 20+ years, for which the amplitude/phase test suffers from gaps in the coverage.
However, even for this frequency the amplitude variations can be expressed well by a mathematical two-mode model.

If we examine these three close frequency pairs together with other (usually more widely separated) close frequencies in FG Vir,
18 pairs of frequencies with separations closer than 0.10 cycle d$^{-1}$ have been detected. It is shown
that the majority of the pairs occur near the theoretically expected frequencies of radial modes. Mode
identifications are available for only a few modes: the only detected radial mode at 12.15 cycle d$^{-1}$ is part
of a close pair.

It is shown that accidental agreements between the frequencies of excited modes can be ruled out because of the
large number of detected close frequency doublets.

\end{abstract}


\section{INTRODUCTION}

The search for the astrophysical reason of the observed amplitude and phase variability associated with stellar
pulsation concerns many different types of stellar pulsators in vastly different stages of stellar evolution. We refer
to this phenomenon as the Blazhko Effect, which was noticed by Blazhko (1907) in the RR Lyrae
star RW Dra. These variations are quite common in RR Lyrae stars, e.g., Szeidl (1988) mentions an
occurrence of 20 - 30 \% in the RRab stars. We are presently engaged in an extensive programme
to examine the Blazhko Effect in RR Lyrae stars. However, the effect has also been noticed among other types of pulsators. Some
examples are:

(i) Cepheids: The cepheids with very short periods often show long-term amplitude variability.
Probably the best-known example is the star V473 Lyr (HR 7308) with a radial pulsation period of 1.49\,d and
a Blazhko period $\sim$ 1258\,d. The observed amplitude variability (by at least a factor of 15) during the Blazhko cycle could
not be fit by a simple two-frequency beating model alone (Breger 1981). This is also evident from the fact
that the amplitude variations are not symmetric on the increasing and decreasing branches (see also Burki, Mayor
\& Benz 1982). If beating is involved, at least three frequencies are required. Some indications for a triplet in the
Fourier spectrum were seen in the Hipparcos measurements analyzed by Koen (2001).

(ii) Another example is provided by the sdB stars: significant amplitude variations are detected with a time scale of about
a year (Kilkenny et al. 1999) or even shorter (Pereira \& Lopes 2004). O'Toole et al. (2002) have argued
that the variability may be caused by the beating of close frequencies with  a frequency separation
of 0.1 cycle d$^{-1}$ or less. However, more detailed investigations may be needed.

(iii) Most $\beta$~Cephei variables also show amplitude variations with a variety of time-scales of
under a year (e.g., in EN Lac, Lehmann et al. 2001) up to decades or even several centuries (see Chapellier 1986).

(iv) White Dwarfs: Handler et al. (2003) analyzed new time-series photometry of two pulsating DB white
dwarfs and concluded that the observed amplitude and period variability cannot be explained by the
beating of multiple pulsation modes.

(v) Many $\delta$ Scuti variables show amplitude variability associated with the multimode pulsation. Similar to
the situation seen in white dwarfs, the same star may appear to change its pulsation spectrum to appear as if they
were different stars at different times. However, detailed analyses of 4 CVn showed that the modes do not disappear,
but are still present at small amplitudes. For the majority of the well-studied stars, frequency pairs with frequency
separations less than 0.06 cycle d$^{-1}$ can be seen in the power spectra. These pairs may be
a reflection of close frequencies, amplitude variability of a single mode or observational errors.
For the star BI CMi, Breger \& Bischof (2002) showed from detailed phase-amplitude tests that
the amplitude variability was the result of beating of separate modes with close frequencies. This paper also
summarizes the situation in these stars.

The Blazhko Effect, therefore, exists in many types of stellar pulsators. It is presently uncertain whether
a single physical mechanism is responsible for the phenomenon in all these stars. We wish to distinguish between two
explanations for the observed amplitude and phase variability observed in so many types of pulsating stars:

(i) The amplitude variability is caused by beating between two (or more) close frequencies. Since amplitude
variability is common, the occurrence of close frequencies cannot be an accidental coincidence between randomly
distributed frequencies. In this hypothesis, a physical mechanism must exist to excite the close eigenfrequencies
to observable levels.

(ii) True amplitude and phase variations not caused by beating. These include periodic behavior, possibly
caused by rotational modulation or magnetic fields, or cycle-to-cycle variations as observed in Miras.

The observational tests to distinguish between the different physical explanations require a very large amount
of data covering many months over several years, so that detailed analyses are so far available only for BI CMi.
In this paper, we want to examine the $\delta$ Scuti star FG~Vir, for which extensive data
and multifrequency analyses are available (Breger et al. 2004, 2005: note that the table of frequencies
given in the latter reference already refers to the analysis carried out in this paper, and lists some preliminary
results). The previous work has shown that in FG Vir three modes seem to possess strongly variable amplitudes.

The present paper also examines the systematics of the occurrence of close frequencies, defined here as multiple frequencies
separated by 0.1 cycle d$^{-1}$ or less.

\section{Method to separate true amplitude variability from beating}

\begin{figure*}
\centering
\includegraphics*[bb=68 294 768 755,width=175mm,clip]{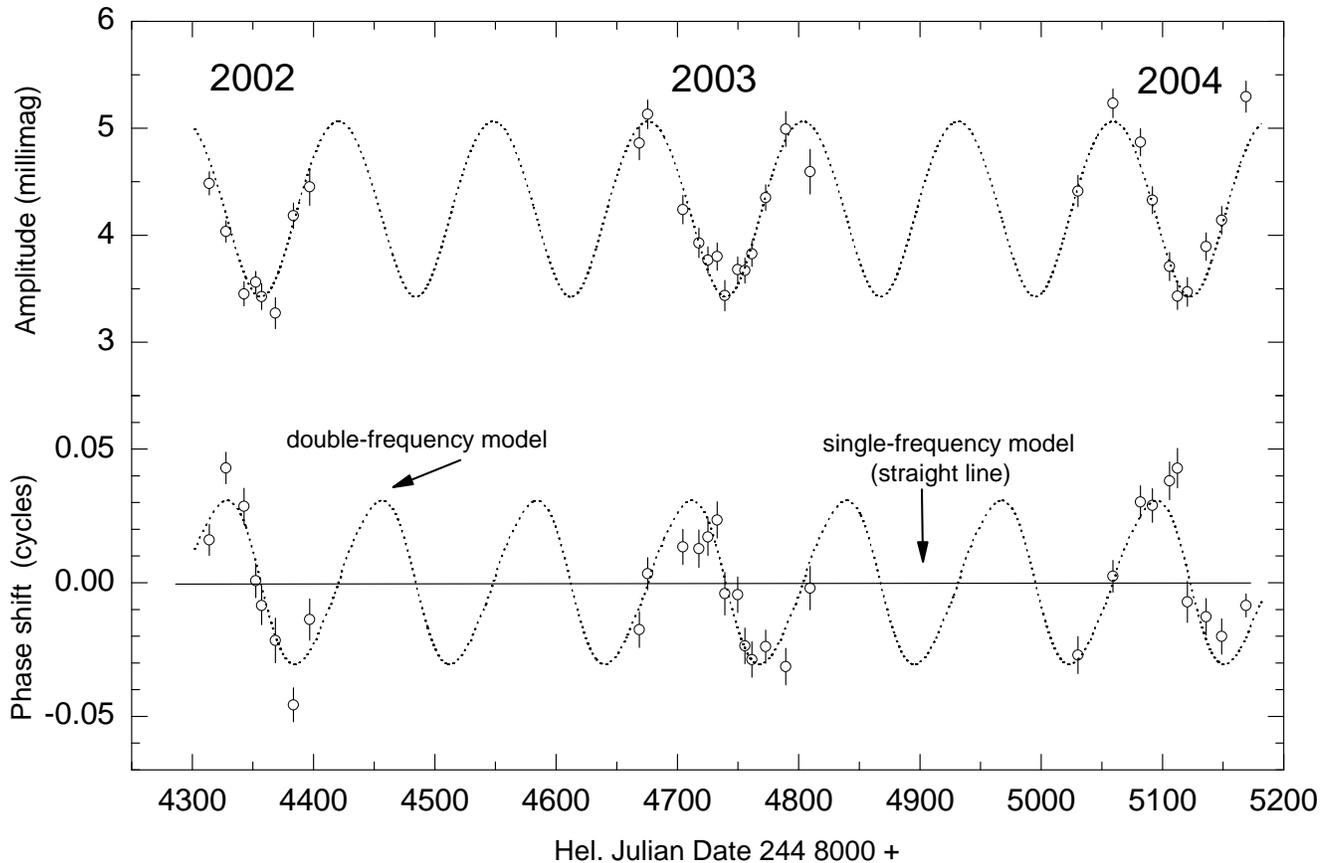}
\caption{Amplitude and phase variations of the 12.15 cycle d$^{-1}$ frequency. The open circles
represent the observations, while the dotted line is the two-frequency fit. Note that a single-frequency model
with only amplitude variability excludes phase changes. The diagram shows that beating between
two frequencies separated by 0.0078 cycle d$^{-1}$ and an amplitude ratio of 0.19 is responsible. This is shown by the
excellent two-frequency fit as well as the correct signature
of beating: the amplitude and phase vary synchronously with a phase shift close to 90$\degr$.}
\end{figure*}

In the Fourier spectrum of short data sets, two close frequencies beating with each other appear as a single frequency with variable
amplitude. For long data sets with sufficient frequency resolution, two peaks will be shown.
In the case of a single frequency with variable amplitude and/or phasing, also two (or more)
peaks occur. Consequently, {\em the appearance of two peaks does not prove the existence of two
separate modes}, while the absence of a double peak does not exclude the possibility. This is a fact well recognized in the literature.

It is important to apply an additional method beyond recognizing two peaks in the power spectrum
to separate the two hypotheses. This is especially the case whenever the detection of the two peaks could
be due to imperfect data coverage such as data length being of the order of the beat frequency.
Fortunately, beating between two (or more) frequencies produces amplitude
and phase variations which are mathematically related. The most extreme and easily recognizable
situation occurs when two close frequencies have the same amplitude: this leads to a half a cycle phase shift at the time
of minimum amplitude of the visible (single) frequency. Even when the amplitudes differ, beating has a specific signature in the amplitude
of phase shifts of an assumed single frequency. In particular, the amplitude and phase vary synchronously
with a phase shift close to 90$\degr$ (Tsesevich 1975).

Therefore, it is possible to separate beating from true amplitude variability by studying the relationship between
the amplitude and phase variations of an assumed single frequency.
The method consists of obtaining an extensive photometric data base, subdividing the data into short time bins and calculating the
amplitude and phase of an assumed optimum single frequency for each time bin. Furthermore, if other pulsation frequencies
are also present,  they need to be corrected for. In practice, large
differences in the amplitudes of the frequencies beating with each other, observational noise, as well as short
(relative to the beat period) data sets may make the recognition difficult.

\section{Data analysis}

The recent multifrequency analysis of FG Vir (Breger et al. 2005) led to the
detection of more than 75 frequencies, of which the vast majority are independent
pulsation modes, rather than harmonics or combination frequencies. It had
already been known before the extensive 2002--2004 data set became available that
some of the frequencies show amplitude variability from year to year. This was
confirmed by the new data set. Furthermore, the discovery meant that FG~Vir
behaved like other well-studied $\delta$ Scuti stars without supplying
an astrophysical explanation for this behavior.

However, the 2002--2004 data set is unusual in that it provides excellent frequency
resolution due to its coverage lasting 98, 160, 165 nights in 2002, 2003 and 2004, respectively.
The coverage is sufficient to relate the amplitude and phasing variability within each year. The
data are ideal to examine Blazhko cycles with periods from days (due to the dense coverage) to several months.

For longer cycles, even the excellent coverage for three years would be insufficient and we have to
include some data from 1992 -- 1996 (1992: Mantegazza, Poretti \& Bossi 1994; 1993:
Breger et al. 1995; 1995: Breger et al. 1998, 1996: Viskum et al. 1998). These older data are
considerably less extensive and the prewhitening of a 75+ frequency solution determined from
2002--2004 would be
problematical. Furthermore, the long interval between 1996 and 2002 is difficult to bridge.
Consequently, the data are used only to look for very long Blazhko cycles. The slightly more extensive
1995 data can be used to check the 2002--2004 results.

The lengthy coverage in the years 2002--2004 makes the data suitable for subdivisions into time bins of one or two weeks.
This allows us to examine the change in pulsational behaviour on a time scale of weeks or months. The obvious
difficulty in this approach lies in the fact that 75+ frequencies of pulsation have been identified for FG~Vir and
that independent one-week solutions including all these frequencies
would be completely unreliable: aliasing as well as the lack of frequency resolution
within the week introduce large uncertainties. It is necessary to remove all the periodic content outside
the frequency range of interest without affecting the modes to be examined for duality: global prewhitening of all
frequencies outside the range of interest is required.

The following approach was used for each frequency region of interest (e.g., 12.1 to 12.2 cycle d$^{-1}$):

(a) For each of the $v$ and $y$ passbands, we calculated an optimum multifrequency fit for the combined 2002--2004 data.
This included all 80 frequencies listed in the table by Breger et al. (2005). This led to 80 amplitude and 80 phase (epoch) values.
These were used to prewhiten the data for all the frequencies outside the chosen frequency range, i.e., fewer than 80 frequencies
were used.

(b) The data were divided into time bins covering approximately one week. The actual time covered in each bin depended on the
coverage and the number of measurements available so that in some cases slightly longer time bins were chosen.

(c) To improve the signal/noise ratio it is useful to analyze the $v$ and $y$ data together. The pulsations of the
star are, however, wavelength dependent. The prewhitened frequencies were treated correctly, since
the prewhitening was performed with separate solutions for the two passbands. This only leaves potential
problems for the mode(s) inside the chosen frequency region of interest. They were solved
in the following manner: The larger pulsation amplitudes
in the $v$ passband were corrected for by scaling the magnitudes by an experimentally determined factor of 0.70 and increasing the
weight in all the calculations by 1/0.70. The phase lags between the two passbands were taken into account by computing for each time bin
the deviations in phase from an overall average for the passband.

\begin{figure*}
\centering
\includegraphics*[bb=68 294 768 746,width=175mm,clip]{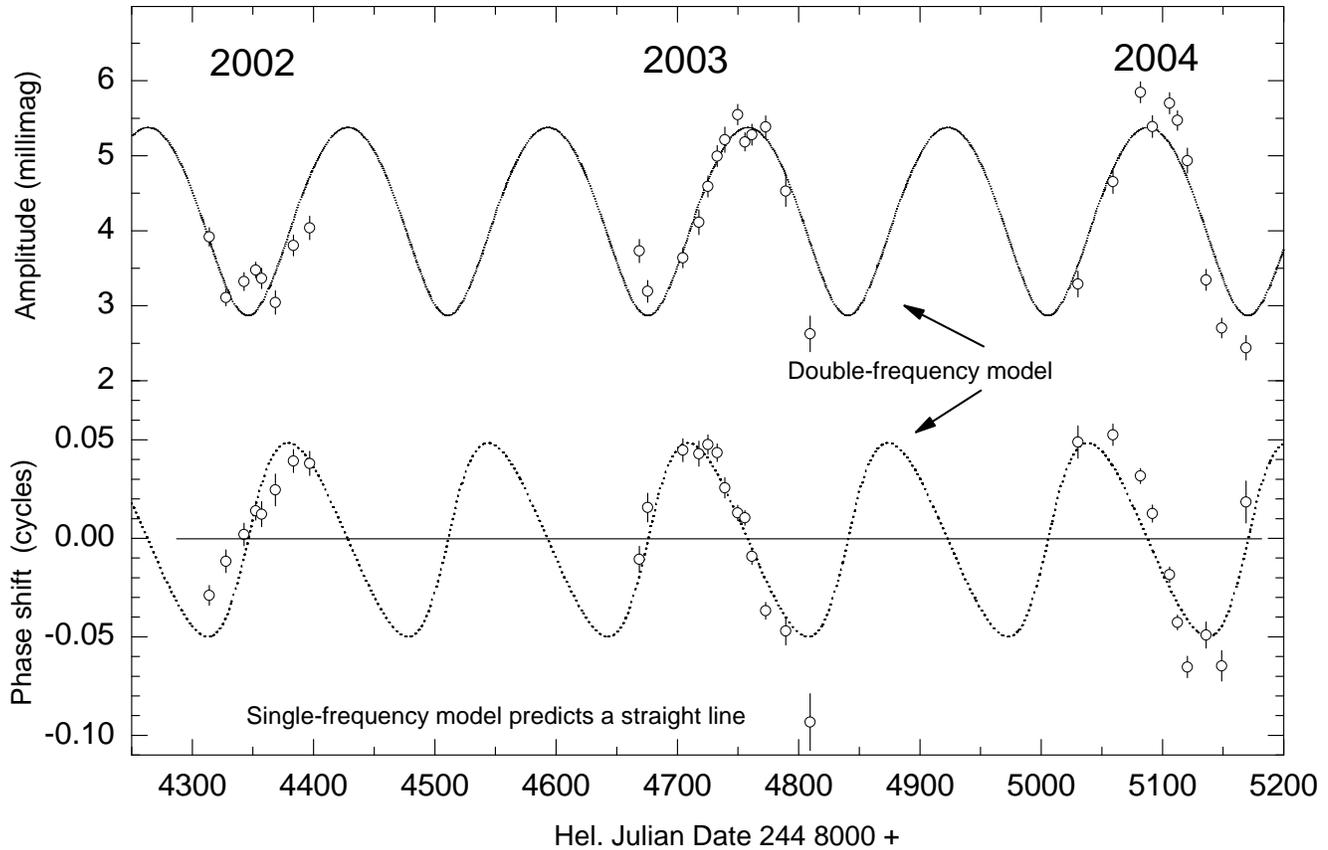}
\caption{Amplitude and phase variations of the 23.40 cycle d$^{-1}$ frequency. The diagram shows that
the variations can be explained by two independent frequencies separated
by 0.0061 cycle d$^{-1}$ and an amplitude ratio of 0.30 (dotted curve). The small deviations are systematic
and can be matched by a third close frequency.}
\end{figure*}

(d) We then adopted the best value of a single frequency inside the frequency region of interest
and computed the amplitude changes and phase shifts for each time bin. The results were examined
for consistency between the separate years and the relationship between the measured amplitude
and phase changes (e.g., see the comparison shown in Fig. 1).

(e) A two-frequency model to give the lowest residuals in magnitudes was determined from Fourier analyses
of the observed light curves and multifrequency fits from PERIOD04 (Lenz \& Breger 2005).
The result was checked by a Fourier analysis of the observed amplitude variations.

(f) The two-frequency model was used to replace the observed magnitudes by predictions for the same times of measurement. We used
the method outlined previously in (d) to compute the single-frequency amplitude changes and phase shifts for each time bin (this time, however,
for the predicted two-frequency data). These were then compared with the values derived from the actual observations and are shown as
dotted lines in Figs. 1 to 4.

\section{Close modes at 12.15 cycle d$^{-1}$}

We have subdivided the available data from 2002 -- 2004 and applied the method described in the
previous section.  The best single frequency was found to be 12.15412 cycle d$^{-1}$. Fig. 1 shows the variations in
amplitude and phase for the three years. In this figure, the uncertainties were calculated by using the
standard relations for calculating amplitude and phasing uncertainties (Breger et al. 1999b)
together with error propogation formulae to account for the measurements in two different passbands.

The variations are similar in all three years with a beat period of 128\,d.
Another important result is that the phase changes are coupled to the amplitude
changes. In particular, minimum amplitude occurs at the time of 'average' phase and
the time of most rapid phase change. This suggests beating by two separate close frequencies.
The visual result is confirmed by a two-frequency model
with the optimum parameters of frequency, amplitude and phase determined by PERIOD04.
 An excellent agreement is obtained for both the amplitude
and phase changes with the two-frequency model. If we extend the analysis to include the 1995 data, we also find an excellent
agreement between the predicted and observed amplitudes and phases of the earlier measurements.

We conclude that the amplitude and phase variations of the 12.15 cycle d$^{-1}$ frequency can be
explained by the beating of two close frequencies.

\section{Close modes at 23.40 cycle d$^{-1}$}

\begin{figure*}
\centering
\includegraphics*[bb=63 407 768 746,width=175mm,clip]{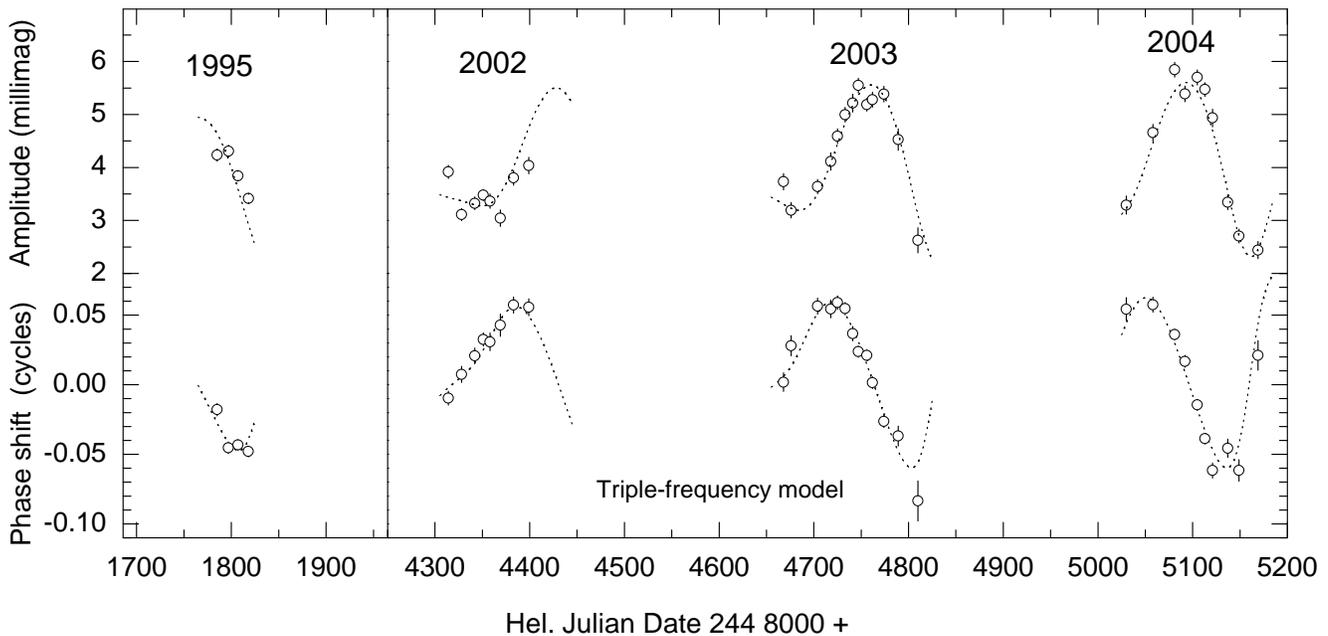}
\caption{Amplitude and phase variations of the 23.40 cycle d$^{-1}$ frequency adding a third frequency
with a small amplitude. We note the improved fit, including that of the 1995 data. See the text for a
short discussion whether or not the star really has a closely spaced triplet.}
\end{figure*}

This frequency also shows considerable amplitude variations from 2 to 6 mmag as well as phase changes.
These variations take place from year to year as well as within a single observing season. This is
shown in Fig.~2. We notice that the situation is quite similar to that of the 12.15 cycle d$^{-1}$ frequency, except
that the Blazhko period is longer by about a month. Again, the signature of beating is evident:
minimum amplitude occurs at the time of 'average' phase and the time of rapid phase change.

The fit of the two-frequency model is quite good, but two items need to be discussed: the amplitude may
not be constant from year to year and one point fits poorly: this point at HJD 245 2809 has by far the
smallest number of measurements because they were obtained at the end of the observing season.
Consequently, the error bars are larger, but the point still deviates by more than 2 standard deviations.

The amplitude increases from 2002 to 2004. This is also seen in a different way:
after prewhitening two frequencies (23.4034 and 23.3973 cycle d$^{-1}$) with constant
amplitudes, the Fourier analysis reveals a third frequency with a small amplitude (Table 1). The
Fourier analysis shows two possible values for this third frequency: 23.3915 and 23.3942 cycle d$^{-1}$.
The first value forms an almost equidistant triplet with the previously found doublet, while the other value
shows a 2:1 distance ratio in frequency spacing. At this stage, we are not able to judge whether or not this
triple frequency splitting has a physical meaning, but note that it does not correspond to the effects
of annual aliasing, which leads to a smaller splitting of 0.0027 cycle d$^{-1}$. Furthermore,
rotational splitting can also be ruled out since it is a factor of 100 larger ($\sim$ 0.53  cycle d$^{-1}$). Trial
calculations with PERIOD04 show that the value of 23.3942 cycle d$^{-1}$ (rather than 23.3915 cycle d$^{-1}$) leads to lower
residuals in every one of the years with observations and is subsequently adopted.

The third frequency provides an excellent fit, but we cannot disprove the possibility that it is just an artefact of true
amplitude variability of one of the two modes involved in the frequency doublet. The main effect of the third
frequency is to introduce slow amplitude changes.

We conclude that the 23.40 cycle d$^{-1}$ frequency consists of two close frequencies beating with each other. The fit
is improved by the addition of a small-amplitude third frequency, which might not be an independent mode.

\section{The amplitude variation of the 19.87 cycle d$^{-1}$ frequency}

\begin{figure*}
\centering
\includegraphics*[bb=63 407 768 746,width=175mm,clip]{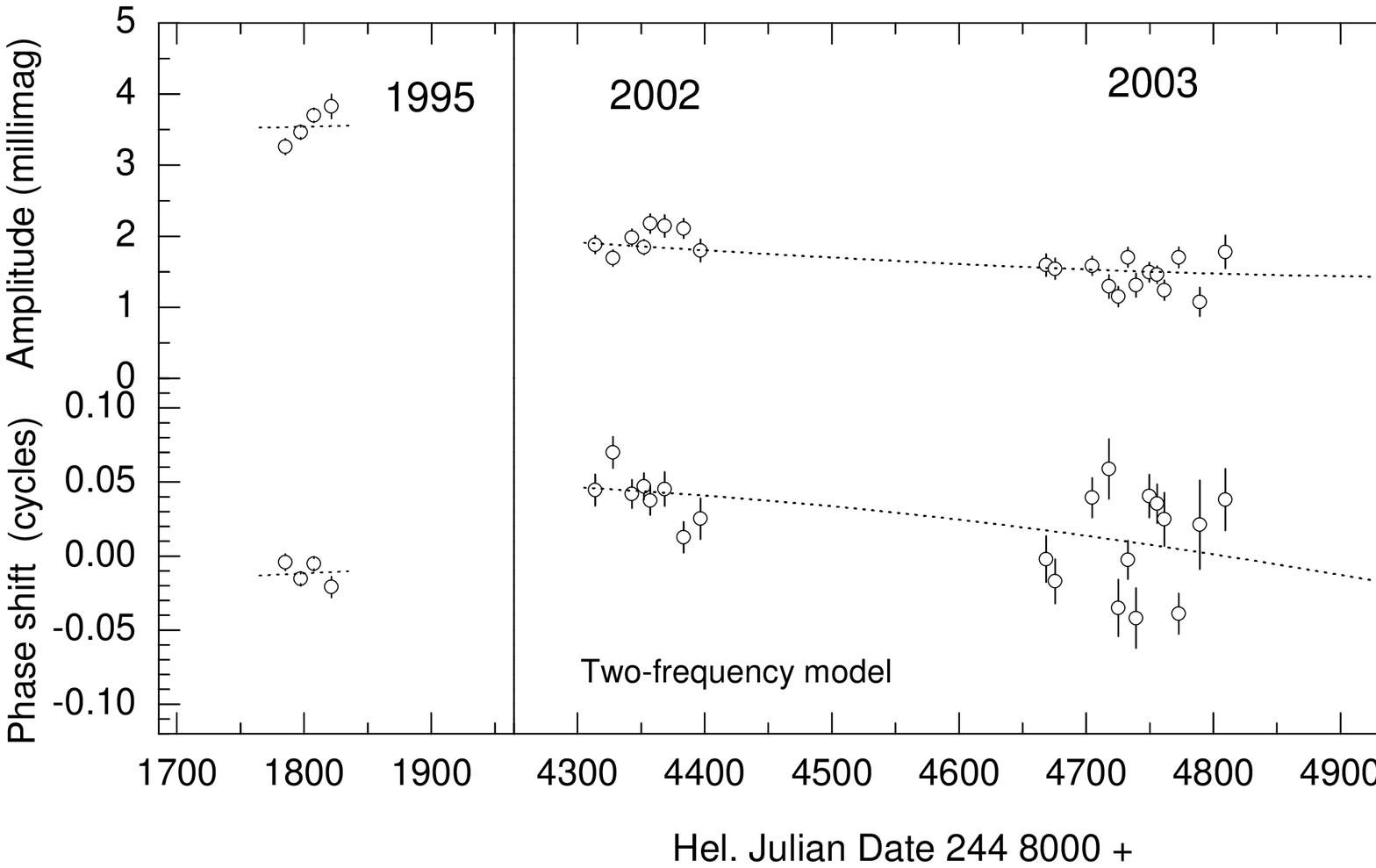}
\caption{Amplitude variations in $y$ of the 19.87 cycle d$^{-1}$ frequency during 1995, 2002, 2003 and 2004.
The open circles represent the observations, while the dotted line is the two-frequency fit.
The diagram shows that the amplitude variability has a long time scale. Furthermore, no single frequency can
fit the observations, even if variable amplitudes are allowed.  The varying sizes of the statistical uncertainties in the observed phases
from year to year are caused by the dependence on the size of the amplitudes.}
\end{figure*}

In the previous papers on FG Vir, the amplitude variations in $V$ of the 19.87 cycle d$^{-1}$ frequency were already
noticed (1985/6: 2.6 mmag, relatively poor data; 1992/3: 4.3 mmag; 1995: 3.5 mmag, 1996: 2.8 mmag; 2002-4: 1.4 to 2.0 mmag).
A cursory inspection of the data also indicates  phase variations. These are the
signatures of the Blazhko Effect. However, the statistical
significance of some of these variations still needs to be examined in a more detailed analysis which
excludes potential errors caused by the other frequencies of the pulsating star.

We consequently apply the technique outlined in the previous section to this frequency as well. We notice
immediately that rapid variations within a single observing season, found above for the other two modes,
are not present. This is demonstrated in Fig. 4, where we have adopted the best single frequency to fit the
1995 and 2002-4 data (19.867908 cycle d$^{-1}$). Both the amplitudes and phases show slow variations on a time scale of many
years.  Note that a frequency value near 19.867801 cycle d$^{-1}$ can fit the 2002-4 phases, but produces
large phase shifts for the 1995 (and for the less extensive 1992, 1993 and 1996 data as well).

The observed variations can be successfully modelled with two close frequencies separated by 0.00013 cycle d$^{-1}$
(Blazhko period $\sim$ 21 years). The good fit is shown in the figure. However, the question examined in
this paper is whether the data suggest that two frequencies are really excited in the star. For such a long
Blazhko period the question cannot be answered in a definite way due to the incomplete Blazhko phase coverage and lack
of observed repeated Blazhko cycles. Consequently, the two-frequency fit might just be no more (and no
less) than an excellent mathematical representation of the observed amplitude and phase variations by
two frequencies. Note that at 12.15 and 23.40  cycle d$^{-1}$ excellent mathematical two-frequency fits were
also available, but that the detailed observed phasing between the amplitude and phase changes provided an additional
excellent verification of the beating hypothesis. This additional verification is weaker for 19.87  cycle d$^{-1}$.

The results for the three tested regions with closest frequencies are summarized in Table~1
together with those for other close frequencies of potential interest.

\begin{table*}
\caption[]{Selected close frequencies detected in the FG Vir data}
\begin{flushleft}
\begin{tabular}{cccccl}
\hline
\noalign{\smallskip}
Modes near & Solution & Frequency & Beat period & Amplitudes in $y$ & Comments\\
cycle d$^{-1}$ & & cycle d$^{-1}$ & & mmag \\
\noalign{\smallskip}
\hline
\noalign{\smallskip}
\multicolumn{6}{l}{Extremely close frequencies to which amplitude-phase tests have been applied}\\
\noalign{\smallskip}
12.15 & 2-frequency &12.15412 & 128 days & 4.21 & 2002--2004 data, 1995 fits well\\
& &12.16195 & & 0.83 \\
\\
23.40 & 2-frequency & 23.40335 & 165 days & 4.12 & 2002--2004 data\\
& & 23.39729 & & 1.25 \\
\\
& 3-frequency & 23.40337 & 167 days & 3.99 & 1995, 2002--2004 data\\
& & 23.39737 & & 1.21\\
& & 23.39426 & 322 days & 0.57\\
\\
19.87 & 2-frequency & 19.867916 & 21 years & 2.60 & 1995, 2002--2004 data\\
& &19.868044 & & 1.16 \\

\noalign{\smallskip}
\multicolumn{6}{l}{Other close frequencies}\\
\noalign{\smallskip}
16.08 & & 16.0711 &  50 days & 1.07 & 2002--2004 data\\
      & & 16.0909 &          & 0.39\\
\\
20.29 & & 20.2878 & 213 days & 1.45 & 2002--2004 data\\
      & & 20.2925 &          & 0.39\\
\\
23.42 & & 23.4034 &  45 days & 4.02 & 2002--2004 data\\
      & & 23.4258 &          & 0.28\\
      & & 23.4389 &  76 days & 0.52\\
\\
24.20 & & 24.1940 &  29 days & 1.58 & 2002--2004 data\\
      & & 24.2280 &          & 4.20\\
\\
26.90 & & 26.8929 &  61 days & 0.27 & 2002--2004 data\\
      & & 26.9094 &          & 0.19\\
\\
34.12 & & 34.1151 & 244 days & 0.49 & 2002--2004 data\\
      & & 34.1192 &          & 0.25\\
\noalign{\smallskip}
\hline
\end{tabular}
\newline Note that some of the values listed differ slightly from those given in Breger et al. (2005). This is due to the fact that here the
$v$ and $y$ colors were analyzed together and that the data used (see Comments) may be different.
\end{flushleft}
\end{table*}

\section{Affinity for radial modes?}

In the previous sections we have examined the observed amplitude variability.
We have shown that at least for two modes this is caused by
beating between close frequencies.

We shall now turn to the observed ubiquity of close frequencies in FG Vir, using the definition of close frequencies as those with a
separation less than 0.1 cycle d$^{-1}$. The majority of these pairs are so well separated
in frequency that these pairs do not show up in data as
single frequencies with variable amplitudes. Consequently, amplitude/phase
tests are not necessary or possible in order to exclude the possibility of spurious results from amplitude variability.
The frequency distribution of close frequency pairs may contain important information about their
physical origin. Fig.\,5 shows that the frequency pairs are not randomly distributed.

In this figure we have plotted the frequency values for all pairs with a separation less than 0.1 cycle d$^{-1}$.
The y-axis represents the frequency separation
for each pair, and each point corresponds to two very close frequencies.
Also shown are the computed frequencies of the radial modes for the FG~Vir model
which fits both empirically estimated global stellar parameters and radial fundamental mode frequency
at the observed frequency 12.154 cycle d$^{-1}$. The observed mode was certainly identified
as radial by Viskum et al. (1998), Breger et al. (2005) and Daszy\'nska-Daszkiewicz et al. (2005).
One additional observed mode, 16.071 cycle d$^{-1}$, was also identified by
Daszy\'nska-Daskiewicz et al. (2005) as a radial mode.
An observed close frequency pair occurs here too (16.071 and 16.091 cycle d$^{-1}$).
However, the frequency values predicted by our model are slightly different (see values below) and
this mode identification needs to be confirmed independently with different methods.

The effective temperature of FG Vir has been determined from
atmospheric models and Stromgren photometry. The luminosity has been determined from
the Hipparcos parallax. The following values have been obtained: $\log T_{\rm eff}=3.869\pm 0.012$,
$\log L/L_{\odot}=1.170\pm 0.055$, in agreement with the values of Daszy\'nska-Daszkiewicz et al. (2005).
The equatorial rotational velocity of FG Vir is found to be $66\pm 16$ km s$^{-1}$ (Zima 2005).

The evolutionary FG Vir model, whose radial
frequencies are given in Fig.\,5, has the following basic parameters:
$M=1.80 M_{\odot}$, $\log T_{\rm eff}= 3.8658$, $\log~L/L_{\odot}=1.120$,
$\log~g=3.980$ and $V_{\rm rot} = 62.5$~km s$^{-1}$ ($V_{\rm rot}{\rm (ZAMS)} = 70$~km/s).
The computations were performed starting with chemically uniform models on the ZAMS,
assuming an initial hydrogen abundance $X=0.70$ and heavy element abundance
$Z=0.02$. The proportions in the abundances of the elements heavier than helium were adopted
according to Grevesse \& Noels (1993). For the opacities, we used the OPAL data
(Iglesias \& Rogers 1996) supplemented with the low--temperature data
of Alexander \& Ferguson (1994). The newest version of the OPAL equation of state
was used (Rogers \& Nayfonov 2002). No overshooting from the convective core was allowed.
In the stellar envelope, the standard mixing-length theory of convection
with a mixing-length parameter $\alpha$ = 0.5 was used. This relatively low value
of the mixing-length parameter was chosen taking into account the conclusion of
Daszy\'nska-Daszkiewicz et al. (2005) that convection is rather inefficient in FG Vir.

We assumed uniform (solid-body) stellar rotation and conservation of global
angular momentum during evolution from the ZAMS.
These assumptions were chosen for simplicity reasons.
The influence of rotation on the evolutionary tracks of $\delta$~Scuti
models was demonstrated by Breger \& Pamyatnykh (1998).
At relatively low rotational velocities of about 60 to 70 km/s,
the evolutionary tracks are located close to those for non-rotating
stellar models. The main effect of rotation is the splitting
of the nonradial mode frequencies. Even for slowly rotating stars,
this splitting is asymmetric due to the second-order effect of rotation.
The whole frequency spectrum of unstable nonradial modes
will be considered in the next paper where possibilities to explain
the appearance of very close frequencies will be discussed. Here we note the
small, but not negligible effects of rotation on the frequency spectrum of radial modes.
This is caused mainly by the small change in the model structure due to the variations
of the effective gravity inside the star. For example, if we fit the frequency of the radial
fundamental mode to the observed value of 12.154 cycle d$^{-1}$, we obtain the frequency
of the 6th radial overtone equal to 34.536 cycle d$^{-1}$ and 34.345 cycle d$^{-1}$ for the equatorial
rotational velocities of 49 km s$^{-1}$ and 81 km s$^{-1}$, respectively. Additional tests show
that a small decrease of the heavy  element abundance (say, to $Z=0.015$) and
a small change of opacities
(by using new OP data according to Badnell~et~al.\,2005 and Seaton~2005) may
result in a very good fit of the frequency of the 6th overtone to
the observed frequency of the close pair at 34.12 cycle d$^{-1}$. Another possibility
to achieve such a fit is to suggest still faster rotation
($V_{\rm rot}$ of about 90-100 km s$^{-1}$) which is outside the allowed limits.
\footnote{In the paper of Breger et al.\,(1999a) a good fit of the frequency
of the 6th radial overtone to the observed frequency at 34.12 cycle d$^{-1}$ was achieved
with the OP opacity data for an FG Vir model of standard chemical composition and
relatively low rotational velocity of 45 km s$^{-1}$ (see Fig.\,7 in that paper).
However, there we used a previous version of the OP data which did not include
important atomic effects in deep stellar interiors. The new OP data are in better
agreement with the OPAL data if we use the same chemical composition for both sets.}

We also tested many other models with the global parameters inside the error box
of the effective temperature and luminosity. Again, we have considered only those models
which fit the radial fundamental mode frequency to the observed frequency of 12.154 cycle d$^{-1}$.
Therefore, only moderate changes in the model parameters and in the input physics were allowed.
We varied the following parameters: (i) mass, (ii) chemical composition
(by choosing $Z=0.015$, for example, and by changing the proportions in the abundancies of
the elements heavier than helium as were determined very recently by Asplund~et~al.\,2004,~2005
for the Sun), (iii) the opacity data
(by using new OP data according to Badnell~et~al.\,2005 and Seaton~2005), and (iv)
the efficiency of overshooting from the convective core. All these models, when fitted to
the observed frequency of 12.154 cycle d$^{-1}$ as the radial fundamental mode frequency, give very
similar frequencies of radial overtones, which do not differ significantly from the frequencies
shown in Fig.\,5.

The figure shows that the majority of the close frequencies are found near expected radial modes, which
are predicted to have values of 12.155, 15.700, 19.442, 23.195, 26.960, 30.704, 34.469 cycle d$^{-1}$ from the fundamental
to the 6th overtone pulsator. These values are calculated for the model with $V_{\rm rot} = 62.5$~km s$^{-1}$.
As it was noted earlier in this Section, the moderate rotation does not influence the
radial mode frequencies significantly: if we fit the frequency of the radial
fundamental mode to the observed value of 12.154 cycle d$^{-1}$, the frequency
of the 6th radial overtone changes from 34.536 cycle d$^{-1}$ to 34.345 cycle d$^{-1}$
when the equatorial rotational velocity changes from 49 km s$^{-1}$ to 81 km s$^{-1}$.
For lower overtones the effect will be proportionally smaller. Therefore, our result that
the frequency pairs cluster around the radial modes would be valid for
every reasonable value of the rotational velocity.

The agreement suggests that an accumulation of nonradial modes around the radial frequencies may occur.
The situation seems to be similar to that observed in RR Lyrae stars (Olech et al. 1999). Another observational fact
may link the Blazhko Effect in RR Lyrae and $\delta$~Scuti stars:  in most close pairs the components have very
different amplitudes (Table 1). Important differences between two types of stars also exist:
in RR Lyrae stars, the radial modes always dominate
in the frequency spectrum. Also, the frequency spectrum of nonradial modes is much sparser and
mode trapping is less effective in $\delta$ Scuti stars than in RR Lyrae stars.

\begin{figure*}
\centering
\includegraphics*[width=175mm]{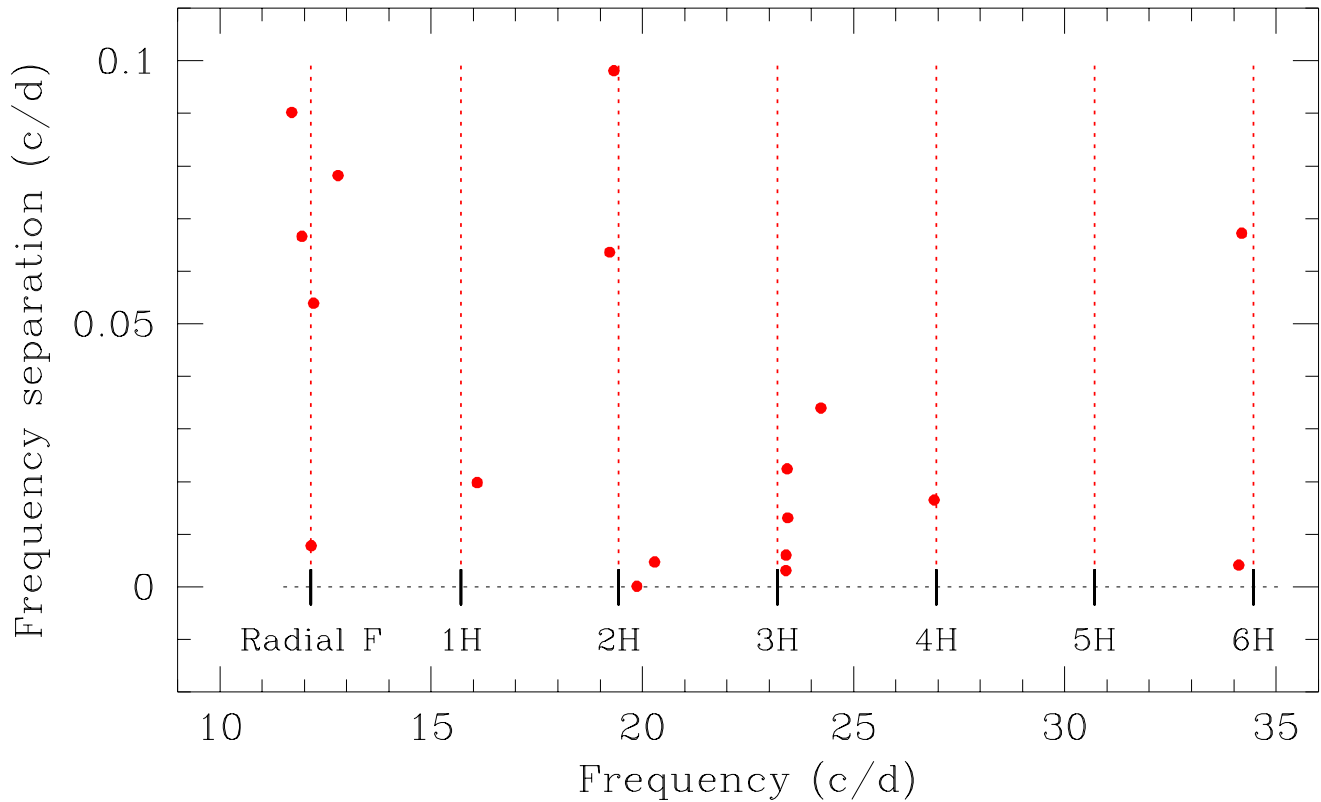}
\caption{Frequencies of close pulsation modes for all pairs with a separation less than 0.1 cycle d$^{-1}$.
The y-axis represents the frequency separations of the pairs. Along the x-axies the frequencies
of the observed close pairs are shown together with the positions of the computed radial modes.
We note that most of the close frequency pairs are situated near
the theoretically predicted frequencies
of the radial modes}
\end{figure*}

\section{Discussion}

FG Vir has such an extensive photometric data coverage
with excellent frequency resolution that the relationship between the
amplitude and phase variations could be examined in detail. The results
presented in the previous sections fully support the interpretation of beating between two (or maybe more) close
frequencies. If we consider the fact that more than 75 frequencies have been detected for
FG~Vir with values between 5.7 and 44.3 cycle d$^{-1}$, the question arises of whether such an
agreement could be accidental. In particular, we need to calculate the probability
of accidental agreements.

Let us consider the 12.15 cycle d$^{-1}$ pair with a separation of 0.0078 cycle d$^{-1}$. If we assume a random
distribution of frequencies and adopt a Poisson distribution, we obtain 0.02 expected pairs.
Of course, the frequencies are not distributed at random. Let us examine a frequency region
which shows the largest number of detected modes, i.e. from 10 to 13 cycle d$^{-1}$ region (Fig.~6).
This increases the predicted number of accidental agreements to 0.03 pairs. For two or more pairs detected in a star,
an explanation of all agreements being caused by accident must be rejected.

\begin{figure*}
\centering
\includegraphics*[bb=22 548 725 752,width=175mm,clip]{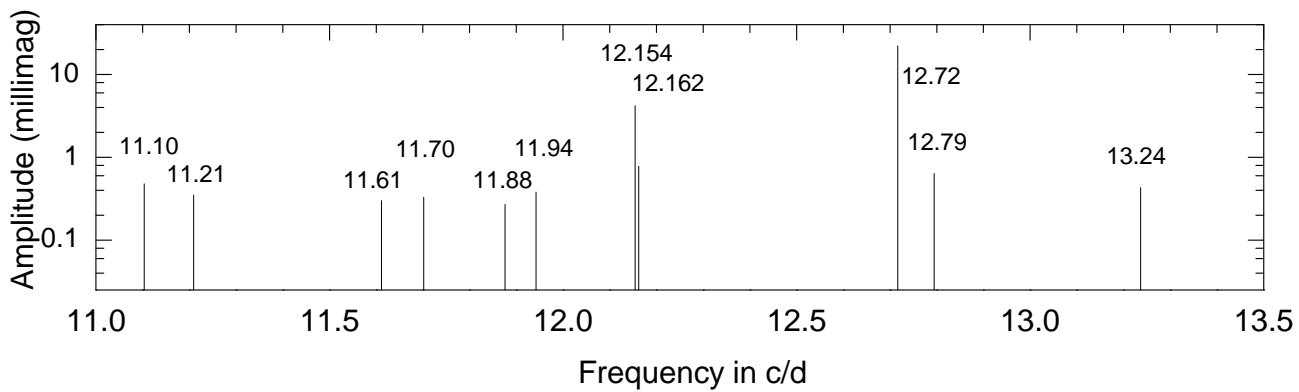}
\caption{Detected modes in the 11 to 13.5 cycle d$^{-1}$ range of FG Vir. The diagrams shows the high
density of modes.}
\end{figure*}

An accidental agreement may exist: the doublet at 42.1030 and 42.1094 cycle d$^{-1}$.
The value of 42.1030 cycle d$^{-1}$ is a 2f harmonic of a relatively high-amplitude (3~mmag) mode at 21.0515 cycle d$^{-1}$. The
42.1030 cycle d$^{-1}$ peak may, therefore, be a consequence of a not completely sinusoidal light curve, rather
than an independent mode.

Fig. 5 has shown the large number of close frequency pairs in FG Vir. There exist additional close frequency pairs in FG Vir beyond those examined
in detail in this paper (e.g., at 34.1151 and 34.1192 cycle d$^{-1}$), but their amplitudes are too small for the tests used in this paper.

We conclude that most of the close frequency pairs found in FG~Vir (and in some other $\delta$~Scuti stars such as BI~CMi) are not accidental
and an astrophysical origin needs to be found. This work is presently in progress.

\section*{Acknowledgements}
It is a pleasure to thank Katrien Kolenberg for many helpful discussions. This investigation has been supported by the
Austrian Fonds zur F\"{o}rderung der wissenschaftlichen Forschung und by the Polish MNiI grant
No.~1~P03D~021~28.

\end{document}